# Classifying and Ranking Microblogging Hashtags with News Categories

Shuangyong Song, Yao Meng
Internet Application Laboratory, Fujitsu R&D Center Co., Ltd.
Beijing 100025, China.
{shuangyong.song, mengyao}@cn.fujitsu.com

*Abstract*—In microblogging, hashtags are used to be topical markers, and they are adopted by users that contribute similar content or express a related idea. However, hashtags are created in a free style and there is no domain category information about them, which make users hard to get access to organized hashtag presentation. In this paper, we propose an approach that classifies hashtags with news categories, and then carry out a domain-sensitive popularity ranking to get hot hashtags in each domain. The proposed approach first trains a domain classification model with news content and news category information, then detects microblogs related to a hashtag to be its representative text, based on which we can classify this hashtag with a domain. Finally, we calculate the domain-sensitive popularity of each hashtag with multiple factors, to get most hotly discussed hashtags in each domain. Preliminary experimental results on a dataset from *Sina Weibo*, one of the largest Chinese microblogging websites, show usefulness of the proposed approach on describing hashtags.

## I. INTRODUCTION

Over the last few years, microblogging is increasingly becoming an important platform for users to acquire information and publish some reviews and personal status. A particular feature of microblogging is the hashtag, a short-hand convention adopted by microblog users to manually assign their posts to a wider corpus of messages on the same topic. They are denoted with the # symbol preceding a short string, often a name or abbreviation [1]. In this way, hashtags are used to represent a topic of the content, which helps tweet search and allows users to join the discussions [2]. However, the arbitrariness of hashtags can lead to mess, making them hard to be showed or searched by domain.

Some existing work have been done to analyze microblogging hashtags. Song et al. took hashtags as topics, and proposed a spatio-temporal model for searching related hashtags in microblogging [3]. Then they further improve this model to detect dynamic association among those hashtags with the time sequence burst technique [4]. Vicient and Moreno presented a novel unsupervised domain-independent methodology to realize the semantic clustering of hashtags, which permits to detect related topics to a given set of tweets [5]. However, those works didn't consider the analysis on the domain classification of microblogging hashtags, which is very important for describing hashtags and helping users more easily get useful information from hashtag based search or browse. In this paper, we propose a new approach for classifying and ranking hashtags by news category information. To the best of our knowledge, no other work on analyzing microblogging hashtags with news categories has been done to date.

News domain information is classified by professional web editor, so it is a well manually tagged classification dataset [6]. For example, on the *SOHU* news website (*news.sohu.com*), news information are classified into 18 categories, such as 'sports', 'social', and 'entertainment'. There are daily more than 1,000 news in each domain on average. In this paper, we first utilize news category information to train domain classification model, and then classify hashtags into different domain with this model. Finally we calculate the domain-sensitive popularity of each hashtag with multiple factors, to get most hotly discussed hashtags in each domain. In the following of this article, we will illustrate some details of our proposed approach, along with some preliminary experimental results.

## II. PROPOSED APPROACH

The proposed approach contains three modules: training of domain classification model, domain classification of hashtags and domain sensitive ranking of hashtags.

### A. Training of Domain Classification Model

We take advantage of news media that contains news articles classified into categories, because there is no information related to categories in microblogging. We gather news articles from *SOHU* news website that classifies the articles into the categories. News articles in each category are gathered together as the classification model training dataset. We extract features of the categories from the news articles using two methods; one is based on a bag-of-words approach and the other is based on a topic modeling approach [6]. We train two different classification models based on bag-of-words features and topic-level features respectively, and then use the average value of two models as our final evaluating indicator. Empirically we choose *Support Vector Machine* (*SVM*) as the training model of our preliminary experiments, more detailed model performance comparison will be considered in the future.

## B. Domain Classification of Hashtags

Hashtag domain distribution is defined as $[d_1, d_2, \ldots, d_n]$, where $d_i$ represents the probability of the $i$-th domain and $n$ is the number of domains. For classifying hashtags, we need to obtain the semantic text of each hashtag. Existing works usually take tweets containing a same hashtag as its semantic text. However, many unrelated spam advertisings tweets are wrongly collected. In this section, we propose a clustering method with self-adaptive threshold to hand this problem, which is simple but effective for filtering spam advertisings. The clustering threshold for a hashtag in our method is defined as 'average Euclidean distance value of all the tweets of this hashtag', then all the tweets with distance smaller than the threshold will be clustered together, finally the tweets in the maximal cluster will be taken as the semantic text of this hashtag. We extract the bag-of-words features and topic-level features of a hashtag respectively from its semantic text, and then get its domain distribution with the two trained domain classification models.

## C. Domain Sensitive Ranking of Hashtags

In this section, we present our method to rank the domain sensitive hot value of hashtags with considering three factors. The first factor we consider for this ranking task is the probability of the hashtag belonging to the classified domain. With the above mentioned domain classification model, we can easily get the probability of a hashtag belonging to each domain. Besides, we consider other two factors. They are the number of the related microblogs of this hashtag, and the posting time of those microblogs. Bigger amount and more recent posting time will make a hashtag hotter. As above discussion, the formula for calculating the domain hot value of a hashtag is given below:

$$H(h_i) = \sum_{j=1}^{N_i} \exp(-\frac{t_p - t_j}{\gamma}) * p(h_i, D_i) \quad (1)$$

where the $h_i$ is a hashtag, and $H(h_i)$ means the hot value of $h_i$, and $N_i$ means the number of microblogs which contain $h_i$, and $t_p$ means the present time, and $t_j$ means the posting time of the $j^{th}$ microblog which contains $h_i$, $(1 \leq j \leq N_i)$. $\gamma$ is the kernel parameter of the temporal decay function, which represents the speed of the decay of users' interest. In this invention, we set it as 7 (days). $D_i$ means the domain which is decided as the domain of $h_i$, and $p(h_i, D_i)$ means the probability of $h_i$ belonging to $D_i$.

## III. PRELIMINARY EXPERIMENTS

We utilize news articles from *SOHU*, and for each domain we collect 10000 articles, which is a subset of the dataset published by *Sogou Labs* [7]. Then we randomly choose 2000 hashtags from our collected *Sina Weibo* dataset with the *public timeline API*[1]. Three graduate students were recruited to annotate the domain classification and domain sensitive ranking results. For each hashtag, every annotator should classify it to a domain and give it an integral score from 1 to 5, where a bigger value means a higher hot value of this hashtag in the belonging domain, based on the annotators personal understanding. For the hashtags which got two or three same classification result, we keep them as the final experimental dataset. In addition, the average kappa score among those three annotators on domain sensitive hot value score is 0.74, which indicates substantial agreements. The classification *Precision* of our model is 0.91, and for the corrected classified hashtags, we got a *NDCG@10* value as 0.898, which shows the usefulness of our model on describing hashtags.

TABLE I. EXAMPLES FOR TOP 5 HASHTAGS IN DOMAIN OF 'SPORTS', 'SOCIAL', 'ENTERTAINMENT' AND 'FINANCE' (TRANSLATED INTO ENGLISH).

| Domains | Entertainment | Social | Sports |
|---|---|---|---|
| Top 5 hashtags | #My_Love_From_The_Star | #MH370_is_missing | #Chinese_Mens_National_Soccer_Team |
| | #Dad_Where_going | #Walking_photograph | #Na_Li_won_the_Australian_Open |
| | #The_Evening_Party_of_the_Spring_Festival | #Law_cast_broadcast | #Evergrande_is_the_champion |
| | #Raymond_Lam | #The_March_1st_terrorist_attacks_in_Yunnan_KunMing_Railway_Station | #_final_match_of_AFC_Champions_League |
| | #Sing_my_song | #public_good_in_microblogging | #_Sochi_Winter_Olympics |

Table 1 gives some examples for top 5 hashtags in domains of 'Entertainment', 'Social' and 'Sports'. For the 2000 randomly chosen hashtags, those three domains got the most proportions, which proves that those domains attract more interest of microblogging users and they can more easily generate topics.

## IV. CONCLUSION AND FUTURE WORK

In this paper, we employed traditional news media and utilized their category information to realize an approach that classifies and ranks microblogging hashtags, which can help users more easily get access to the domain related hot topics. In our future work, we plan to design a model for keyword based hashtag search, and a hashtag ranking mechanism with additionally considering user interest and user influence will be also designed. Besides, daily domain-sensitive hashtag ranking is more useful and it is in the plan of our further experiments.

---
[1] http://open.weibo.com/wiki/2/statuses/public_timeline